# Valley polarized relaxation and upconversion luminescence from Tamm-Plasmon Trion-Polaritons with a MoSe$_2$ monolayer

N. Lundt[1], P. Nagler[2], A. Nalitov[3], S. Klembt[1], M. Wurdack[1], S. Stoll[1], T.H. Harder[1], S. Betzold[1], V. Baumann[1], A.V. Kavokin[3,4], C. Schüller[2], T. Korn[2], S. Höfling[1,5] and C. Schneider[1]

[1]*Technische Physik and Wilhelm-Conrad-Röntgen-Research Center for Complex Material Systems, Universität Würzburg, D-97074 Würzburg, Am Hubland, Germany*

[2]*Department of Physics, University of Regensburg, Regensburg D-93040, Germany*

[3]*Physics and Astronomy School, University of Southampton, Highfield, Southampton, SO171BJ, UK*

[4]*SPIN-CNR, Viale del Politecnico 1, I-00133 Rome, Italy*

[5]*School of Physics and Astronomy, University of St. Andrews, St. Andrews KY 16 9SS, United Kingdom*

**Transition metal dichalcogenides represent an ideal testbed to study excitonic effects, spin-related phenomena and fundamental light-matter coupling in nanoscopic condensed matter systems. In particular, the valley degree of freedom, which is unique to such direct band gap monolayers with broken inversion symmetry, adds fundamental interest in these materials. Here, we implement a Tamm-plasmon structure with an embedded MoSe$_2$ monolayer and study the formation of polaritonic quasi-particles. Strong coupling conditions between the Tamm-mode and the trion resonance of MoSe$_2$ are established, yielding bright luminescence from the polaritonic ground state under non-resonant optical excitation. We demonstrate, that tailoring the electrodynamic environment of the monolayer results in a significantly increased valley polarization. This enhancement can be related to change in recombination dynamics shown in time-resolved photoluminescence measurements. We furthermore observe strong upconversion luminescence from resonantly excited polariton states in the lower polariton branch. This upconverted polariton**

luminescence is shown to preserve the valley polarization of the trion-polariton, which paves the way towards combining spin-valley physics and exciton scattering experiments.

**Introduction**

Two-dimensional atomic crystals of transition metal dichalcogenides (TMDCs) have advanced to become one of the most popular materials to study the physics of excitonic quasiparticles. This is partly a result of the enormous exciton binding energies up to 550 meV[1,2] and of the materials' unique spin-related and non-linear properties[3–5]. The latter are a direct consequence of the broken inversion symmetry of the monolayers in combination with a strong spin-orbit coupling inherited by the transition metal atoms, lifting the degeneracy of the high symmetry k points in the honeycomb lattice[6]. This valley degree of freedom has now been vastly explored in $WS_2$, $WSe_2$, $MoS_2$ and $MoSe_2$[6–10]. Surprisingly, while valley polarization in $WS_2$, $WSe_2$ and $MoS_2$ is routinely observed even under non-resonant excitation conditions up to room temperature[11,12], spin-valley dephasing leads to a rapid depolarization in $MoSe_2$[13,14]. The dynamic process of valley polarization and depolarization strongly depends on the carrier redistribution, scattering and emission lifetime, and thus can be tailored by coupling the excitonic resonances to microcavity modes. In this context, in particular the strong coupling regime is of great interest. Strong coupling leads to a hybridization of light and matter modes and provides a large versatility in tailoring photonic and excitonic properties, such as particle interactions, scattering, and spinor related effects.

Here, we study a system composed of a single $MoSe_2$ monolayer embedded in a III-V – metal hybrid photonic microstructure[15–19]. Substrate engineering allows to transfer a significant part of the oscillator strength form the exciton resonance to the trion resonance. Thus, strong coupling with the $MoSe_2$ trion resonance can be established and is manifested by the characteristic energy-momentum dispersion relations of the polariton branches by angle-resolved photoluminescence measurements. Time-resolved measurements reveal that the recombination dynamics speed up, which in turn results in a higher time-average valley polarization in the polariton ground state as compared to bare monolayers. Furthermore, we detect upconversion luminescence from the upper trion-polariton branch as the system is driven resonantly into the lower polariton ground state. Due to the fast conversion process, the valley polarization is macroscopically preserved in this proof-of-principle experiment, resulting in an elliptical polarization with a circular polarization degree of 13% and a linear polarization of 26 % following the circular pump.

**Sample characterization and methods**

The investigated monolayer flake is shown in Fig. 1a. The monolayer of $MoSe_2$, mechanically exfoliated from a bulk crystal, was transferred with a polymer stamp onto the top GaInP layer (10nm) of a distributed Bragg reflector (DBR), consisting of 30 pairs of AlAs/AlGaAs layers (Fig. 1d). The photoluminescence from this monolayer (Fig. 1b), recorded at 5 K under 532 nm excitation (Nd:YAG laser) is strongly trion-dominated, with linewidths of 9.3 meV and 9.7 meV for exciton ($E_X$ = 1.6681 eV) and trion ($E_{X^-}$ = 1.6334 eV), respectively. It should be noted that both the very high ratio of trion to exciton intensity of 80 and the very high trion dissociation energy of 34.7 meV serve as a strong indicator of a large free electron accumulation in the monolayer[20,21] which is a consequence of the chosen GaInP substrate[22]. If the trion is excited quasi-resonantly via the exciton resonance (743nm, tunable Msquared Solstis Ti:Saphire laser, cw mode) the emission linewidth decreases to 7.6 meV, along with a slight blue-shift and a 20-fold increase in intensity (Fig. 1b inset). Interestingly, after we cap the monolayer with 80 nm poly(methyl methacrylate) (PMMA), part of the oscillator strength is transferred back to the exciton resonance along with a red-shift of the exciton to 1.6632 eV and blue-shift of the trion to 1.6354 eV. The reduction of the trion dissociation energy down to 27.8 meV can be understood as a consequence of the reduced density of free carriers in the system after PMMA capping. In turn, the exciton and trion screening from free carriers is also reduced, which ultimately results in a redshift of the exciton and a blueshift of the trion[21]. Nevertheless, the PL signal is still strongly trion-dominated. A reflectivity measurement of the monolayer confirms that the oscillator strength is actually transferred from the exciton to the trion which is in contrast with previous observations of $MoSe_2$ monolayer on different substrates and other doping mechanisms[20] even after PMMA capping (Fig. 1c). In fact, the integrated area of the absorption dips is a more valid and even proportional measure of the oscillator strength. This observation is of major importance for the coupling of the trion resonance to an optical mode and is also remarkable since conventional photo-doping does not result in a comparable transfer of oscillator strength[20].

The photonic Tamm structure is completed by evaporation 60 nm of gold on top of the PMMA layer (Fig. 1d). The bottom DBR supports a very high reflectivity of 99.97 % in a spectral range between 745 nm and 790 nm. This type of photonic microstructure features a strong field enhancement close to the metallic interface (see Fig. 1e), which has proven to suffice for promoting polariton formation with embedded InGaAs[17], GaAs[18] and II/VI[19] based quantum wells at cryogenic temperatures and more recently, with $WSe_2$ and $MoS_2$ monolayers at ambient conditions[23,24]. The layer thicknesses, illustrated in Fig. 1e as the sequence of the corresponding refractive indices, were designed to promote an optical mode energetically close to the trion resonance and spatially overlapping with the monolayer. The optical mode profile calculated by the transfer matrix method in Fig. 1e exhibits a maximum at the monolayer position. A reflectivity measurement of the empty cavity (next to the monolayer), presented in Fig. 1f, confirms that the photonic structure has a resonance energy $E_c$ of 1.6409 eV close to the trion resonance of $MoSe_2$ with quality factor of 720 (2.28 meV full width half maximum). An exact match between trion and photon mode is very challenging from the experimental point of view, since the PMMA thickness affects the optical resonance energy very sensitively.

The spectra were recorded in a single shot in an angle-resolved setup via imaging the objective's Fourier plane onto a high resolution charge coupled device (CCD) array chip. In order to cover a large emission angle, thus accessing a sizeable spectral tuning range, we use a high magnification (50 x) microscope objective with a numerical aperture of 0.65. Since the polariton in-plane momentum $k_\parallel$ is proportional to $\sin(\theta)$, with $\theta$ being the PL emission angle, this allows us to project an in-plane momentum range of up to 5.5 µm$^{-1}$ onto the CCD chip of our spectrometer in the far-field imaging configuration. For polarization measurements we use a λ/4 waveplate to generate $\sigma^+/\sigma^-$ polarized light and analyzed the emitted signal with a rotatable λ/4 waveplate followed by a linear polarizer. We used a beam splitter preserving 98% polarization. In addition, the incident laser was analyzed by a polarimeter at various positions in the optical path in order to ensure a circular degree of polarization more than 99.9%. A detailed sketch of the setup can be found in supplementary S4.

Time-resolved photoluminescence (PL) measurements were performed in a self-built confocal microscope setup. A frequency-doubled pulsed fiber laser system (TOPTICA TVIS, pulse length (FWHM) 180 fs, pulse repetition rate 80 MHz) tuned to a central wavelength of 700 nm was used as excitation source, coupled into a 100x microscope objective and focused to a spot diameter of less than 1 micron on the sample surface. The PL from the sample was collected using the same objective and coupled into a grating spectrometer, where it was detected using a streak camera coupled to the spectrometer and electronically synchronized with the pulsed laser system. The temporal resolution of this setup (HWHM of the pulsed laser trace) is below 4 ps. The sample was mounted on the cold finger of a He-flow cryostat and positioned beneath the microscope objective with sub-micron resolution using a motorized xy table. All experiments were conducted at 5K.

**Results**

We first probe our device under 743 nm optical excitation (exciton resonance). We observe an intense luminescence signal from two curved branches (Fig. 2a), which follow the dispersion relation of cavity trion polaritons indicated by the dashed lines in Fig. 2a. The line spectra at various in-plane momenta can be fitted and the resulting peak positions were plotted as a function of the corresponding wave vector in Fig. 2b. The extracted data was then fitted by a standard coupled oscillator approach, which allows us to determine a normal mode splitting of 5.2 meV and a positive detuning $\Delta = E_C - E_X$ of 4.2 meV at k=0. Also, the trion energy of 1.6358 eV according to the coupled oscillator fit, is in excellent agreement with the previously measured trion energy under PMMA coverage (1.6354 eV). We note, while the formation of exciton-polaritons with transition metal dichalcogenides has now been observed in different photonic structure and under various conditions[23,25,26], trion polaritons are harder to observe and are subject of vivid discussions. In particular, they currently enjoy a lot of attention resulting from the interest in combining polaritonics with collective fermionic excitation[27–29] (such as superconductivity effects). In addition, we have investigated the valley polarization which manifests in the circular degree of polarization of the photoluminescence. Fig. 2c shows the PL spectra of the lower polariton branch at $k_\parallel = 0$, excited with $\pmb{\sigma}^-$ polarized light and detected in $\pmb{\sigma}^-$ and $\pmb{\sigma}^+$ configuration. This measurement reveals a circular degree of polarization P = 12% in contrast to less than 1% on a bare monolayer[30], demonstrating that tailoring the electrodynamic environment can significantly enhance the valley polarization.

In order to investigate the origin of this effect in more detail, we carried out time-resolved PL measurements on the trion-polariton as well as on a trion resonance of a bare $MoSe_2$ monolayer placed on the same substrate. The fast response of our Streak-Camera setup allows us to capture both the formation (rise time) as well as the relaxation dynamics (decay time) of our investigated states by fitting the time traces with a convolution of an exponential rise, the streak camera response and an exponential decay (see supplementary S6 for more details). The induced population density for the excitation power range used in the experiments is estimated to $10^6$-$10^9$ cm$^{-2}$ (see supplementary S5), which is in a regime where interaction processes of excitons and trions should not be dominant[31]. However, polariton-polariton scattering can be play a role resulting from their larger spatial extension. Fig. 3a depicts the PL time traces of the trion and the trion-polariton, as well as the system response to a laser pulse. It can be seen that the trion-polariton has a significantly longer rise time, while its decay time is much faster than for the trion. More quantitatively, Fig. 3b depicts the trion rise and decay time as a function of the averaged excitation power. Here, the rise time can be interpreted as the trion formation and the average relaxation time into the light cone. The trion formation time slightly decreases with increasing pump power since the trion formation depends on the concentration of excited, free charge carriers. The rise time on the order of 4.5 ps is in good agreement with previous measurements[32] given the different substrates and flake-to-flake variations. The decay time on the order of 25 ps is close to what has been identified as the radiative decay time of trions in $MoSe_2$ monolayers[33] at cryogenic temperatures.

In case of the trion-polariton, rise time and in particular decay time must be interpreted differently, since the radiative lifetime of the polariton ground state is expected to be significantly below the resolution limit due to the light fraction of the polariton. In comparison to the trion, the trion-polariton exhibits a significantly prolonged rise time. This characteristic turn-on delay has been observed for GaAs-polariton samples[34] and partly reflects the relaxation dynamics from reservoir states into polariton states. This relaxation is typically associated with the emission of multiple phonons. However, phonon emission is reduced in polariton modes due to its curved in-plane dispersion. This so-called bottleneck effect is well known from exciton-polaritons in more conventional material systems[35–37]. The initial decrease in rise time with power can again be explained by faster trion-formation at higher concentrations. In addition to phonon-polariton scattering, there is a second scattering mechanism for polariton relaxation, namely polariton-polariton scattering. This mechanism strongly acts on the depopulation of the reservoir states, which manifests in the decay time. Here, the decay time of the trion-polariton signal is significantly reduced to about 12 ps compared to the free trion (by a factor of 2). This relaxation enhancement is well described in the literature[38] and evidences fast relaxation into polariton states facilitated by polariton-polariton particle scattering processes. Both, this distinct enhancement of polariton relaxation and also the assumed very fast radiative decay are highly beneficial to maintain the valley polarization of the injected reservoir, as shown in Fig. 2c).

Since relaxation dynamics into the polaritons ground state and the dynamics of its subsequent decay play an important role to maintain valley polarization, we furthermore test the inverse process: Exciting our system resonantly into the lower polariton energy ground state with a narrow band single frequency laser (energy 1.6336 eV) and observing PL from higher energy states. While the strong resonant laser background prevents us from studying the resonance fluorescence from the ground state, we can monitor a significant population from states with a large in-plane momentum associated with the upper polariton branch. The excitation/emission scheme of this experiment is illustrated in Fig. 4a.

Such a population can be the result of various phenomena: Phonon absorption has been evidenced to cool polariton systems efficiently, followed by anti-stokes luminescence from higher energy states[39]. This process has been shown to be particularly efficient for moderate Q factors and polaritons with significant matter component (as in our case). In addition, polariton luminescence following the resonant absorption of two photons has been demonstrated[40]. Related processes have recently been observed in $WSe_2$[41,42], even under rather weak CW laser excitation. In our experiment, the upconversion energy of about 30 meV can be plausibly explained by both mentioned processes.

To shed more light into the origin of the luminescence in our device, we plot the emitted intensity of the converted emission as a function of the injection power in double logarithmic scale in Fig 4d. While fast and slow phonon cooling processes should yield and approximately linear population of higher states with the injection power, a two-photon absorption process is expected to exhibit a quadratic power dependency. The extracted power coefficient of 1.4475 indicates, that both processes indeed contribute to the luminescence from the upper branch in the chosen excitation regime.

In order to address the trion-polariton valley degree of freedom in the polariton frequency upconversion experiment, we resonantly inject polaritons into the lower polariton branch with a $\sigma^-$ polarized laser. The polarization state of the converted polariton luminescence from the upper branch is then studied by a fourier analysis methode using a rotating $\lambda/4$ waveplate[43]. We carefully map out the state of polarization by tuning a $\lambda/4$ waveplate in front of a linear polarizer, and measure the angle-resolved PL spectra as a function of the $\lambda/4$ waveplate rotation angle. We then plot the integrated peak intensity I versus rotation angle $\theta$. All four Stokes parameters $S_0$, $S_1$, $S_2$ and $S_3$ can now be extracted by fitting the acquired function with

$$I(\theta) = \frac{1}{2}(S_0 + S_1 cos^2(2\theta) + S_2 \sin(2\theta)\cos(2\theta) - S_3 sin(2\theta))$$

and the condition

$$S_0^2 \geq S_1^2 + S_2^2 + S_3^2$$

Whereas $S_0$ describes the total intensity of the optical field, $S_1$ describes the degree of polarization in the horizontal-vertical linear basis, $S_2$ describes the degree of polarization in diagonal linear basis and $S_3$ describes the degree of polarization in circular basis. Alternatively, the degree of circular polarization can be extracted by measuring the intensities in $\sigma^+$ and $\sigma^-$ basis and using $S_3 = \frac{I(\sigma^+)-I(\sigma^-)}{I(\sigma^+)+I(\sigma^-)}$. The corresponding spectra for this case are presented Fig. 4b. In principle, the characteristic oscillation of the emitted intensity has a period of 180°, which is sufficient of extract all Stokes parameters. For completeness, and to test the reproducibility of our experiment, we measure a full rotation of 360° (see Fig. 4c). The Stokes fit results in $S_1$ = 26.5%, $S_2$ = 0.5% and $S_3$ = 13%. The circular polarization degree $S_3$ of 13 % demonstrates that polaritons with the selected valley index have been injected, and more importantly, that significant valley polarization has been preserved in the frequency upconversion process. This clearly points towards the significant role of phonon absorption in the upconversion process, since two photon absorption cannot be expected to preserve the valley index[42]. The linear-horizontal polarization degree $S_3$ of 26.5 % is somewhat surprising, since we did not excite a valley superposition in our experiment (as in the case of linearly polarized excitation). This observation can be interpreted as a consequence of the linear polarization splitting of our microcavity resonance at the measured energy. The linear polarization splitting (TE-TM splitting) amounts to 4.8 meV, acts as an effective magnetic field on our valley polaritons and is sufficient to induce a linear polarization in our signal. A fully quantitative treatment of the underlying effect is discussed in supplementary information S2.

In conclusion, we have observed trion-polaritons in a hybrid III-V and polymer Tamm-plasmon-polariton device featuring an integrated single atomic layer of the transition metal dichalcogenide $MoSe_2$. We demonstrate a transfer of the oscillator strength from the exciton to the trion resonance solely by substrate engineering. Time resolved measurements demonstrate a fast relaxation of trion-polaritons, which result in a significant degree of valley polarization under non-resonant injection in time-averaged PL measurements. In addition, our first measurements of polariton dynamics with TMDC monolayers contain further implication regarding condensation experiments. Macroscopic preservation of the polariton valley polarization is further demonstrated in a resonant frequency upconversion experiment. Our results pave the way towards exploiting valley physics in the strong couping regime. In addition, our results demonstrate a pathway towards optical cooling of polaritons as well as multi-photon abosorption experiments with two dimensional materials in the strong coupling regime.

This work has been supported by the State of Bavaria and the ERC (unlimit-2D) as well as the DFG via grants GRK 1570, KO3612/1-1 and SFB 689

Correspondence and requests for materials should be addressed to Christian Schneider and Nils Lundt (christian.schneider@physik.uni-wuerzburg.de, nils.lundt@physik.uni-wuerzburg.de)

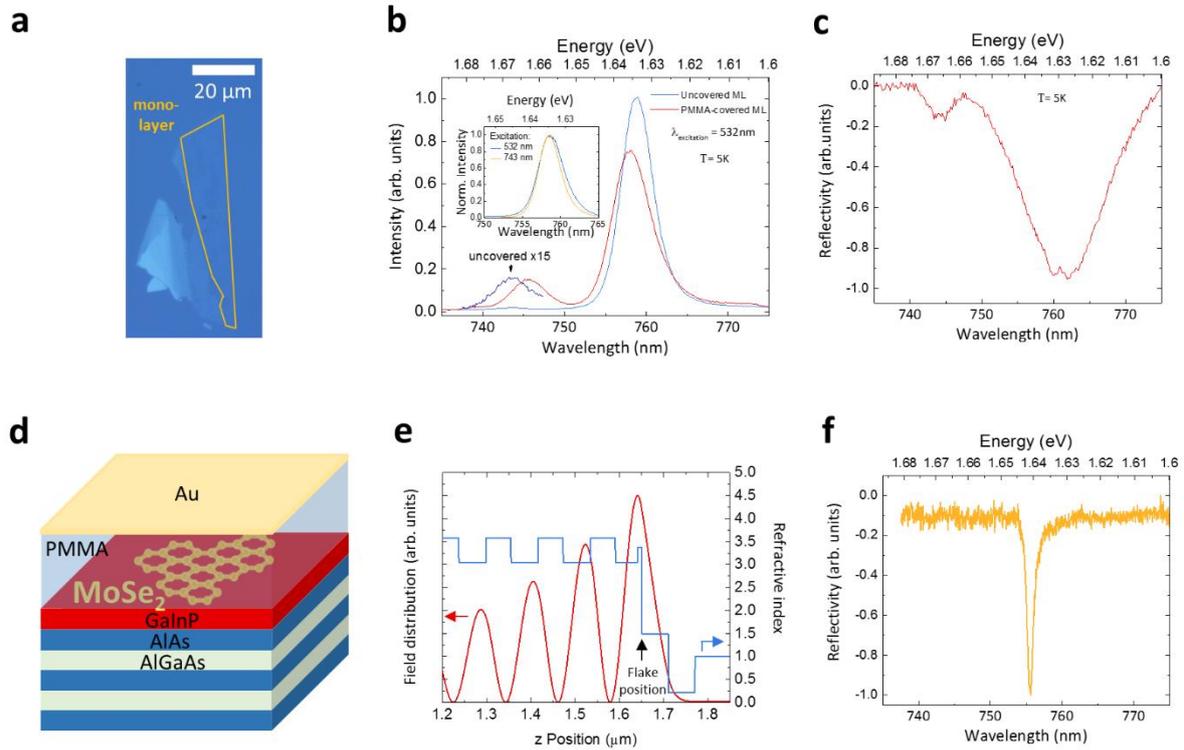

**Figure 1 | Characterization of the investigated monolayer and the photonic Tamm structure:** a) microscope image of the investigated MoSe$_2$ monolayer transferred onto the bottom DBR structure. b) PL spectra of the excitonic and trionic resonances under 532nm excitation at 5 K before and after PMMA capping (blue and read lines, respectively). The inset presents the comparison between non-resonant (532 nm) and close-to-resonant (743 nm) excitation of the trionic resonance. c) White light reflectivity spectrum of both resonances after PMMA capping. d) Schematic illustration of the photonic Tamm structure including the embedded MoSe$_2$ monolayer. e) Layer sequence of the top part of the Tamm structure represented by the corresponding refractive indices (blue profile) and the simulated field distribution within the Tamm structure. f) Reflectivity spectrum of the empty Tamm structure, yielding a Q factor of 720, with a resonance close to the trion absorption.

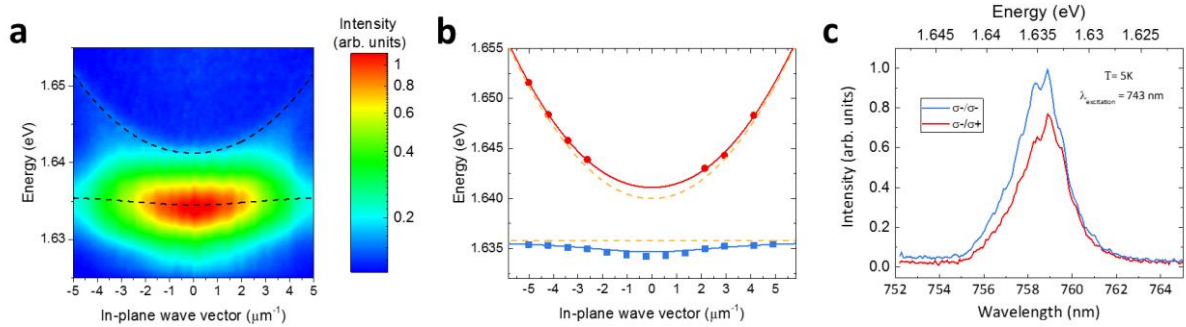

**Figure 2 | Dispersion relation of trion-polaritons:** a) Angle-resolved PL measurement acquired in one shot in a Fourier setup. Lower and upper polariton branch are indicated by the black dashed lines. b) Peak positions acquired from line spectra fits as a function of the in-plane wave vector. Blue and red lines result from a coupled two-oscillator fit, whereas uncoupled trion and photonic mode dispersions are indicated by yellow dashed lines. c) Line spectra of the lower polariton branch at $k_{||}$ = 0 µm$^{-1}$ upon excitation with $\sigma^-$ and detected in $\sigma^-$ and $\sigma^+$ configuration.

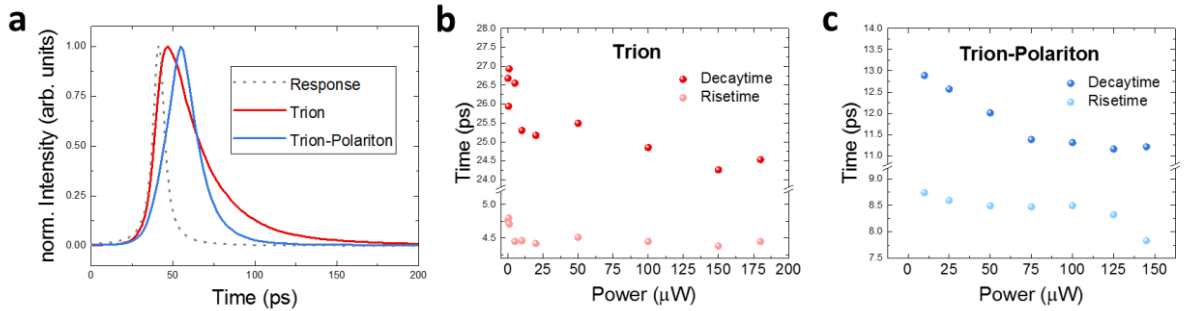

**Figure 3 | Time-resolved photoluminescence measurements:** a) Exemplary time traces of the trion and trion-polariton response. The gray dotted line illustrates the streak-camera response function ($E_{Laser}$ = 1.771 eV). In comparison to the trion ($E_{Trion}$ = 1.6334 eV), the trion-polariton ($E_{Trion\text{-}polariton}$ = 1.6336 eV) exhibits a slower rise, but a faster decay. b) Rise and decay time of the trion as a function of the averaged excitation power. c) Rise and decay time of the trion-polariton as a function of the averaged excitation power.

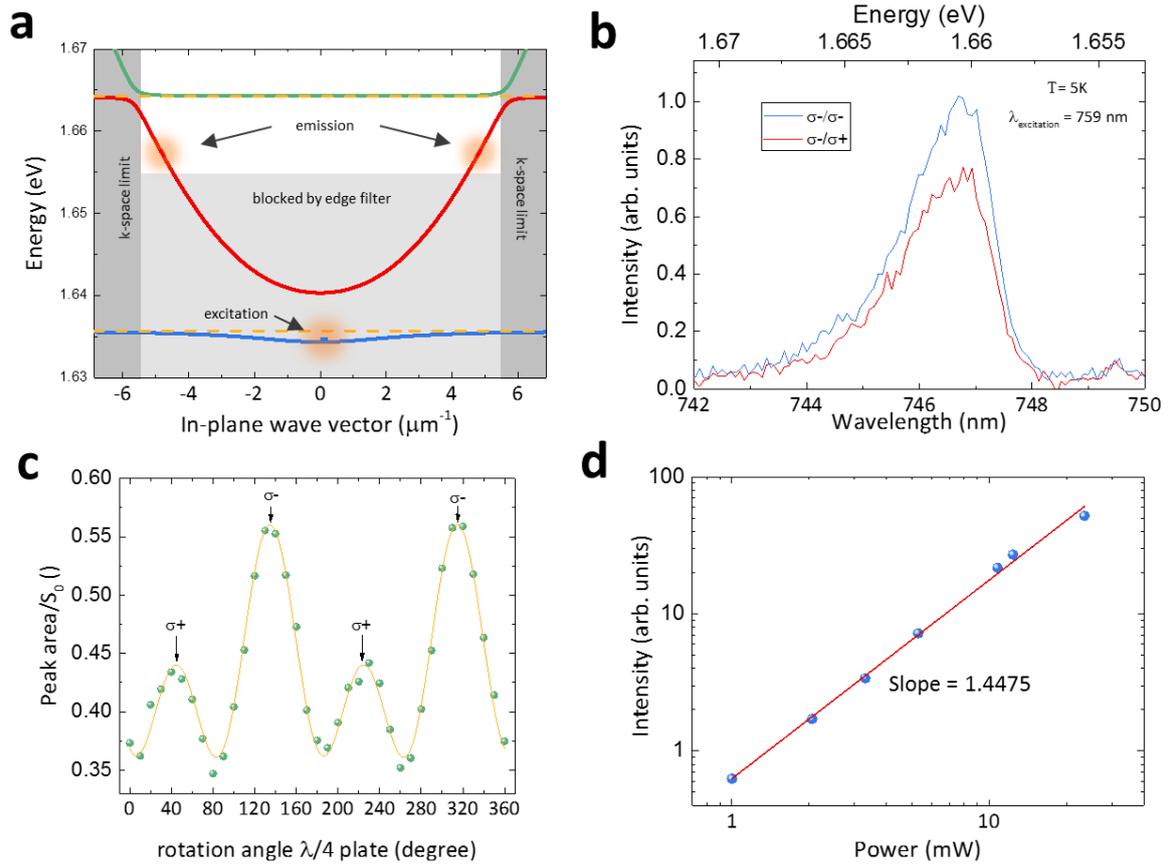

**Figure 4 | Polariton upconversion photoluminescence measurement:** a) Schematic illustration of the excitation/detection scheme in this experiment: The lower trion-polariton branch is resonantly excited, while the emission is detected above 1.655 eV at high in-plane k-vectors. Detection cut-off by an edge filter and k-space limit given by the NA of the objective are indicated as shaded areas. b) Line profiles of the detected upconversion PL under $\sigma^-$ polarized light excitation measured in $\sigma^-$ (blue line) and $\sigma^+$ basis (red line). c) Integrated peak intensity as a function of the rotation angle of the $\lambda/4$ waveplate and the Stokes fit (yellow line). d) Double-logarithmic illustration of an upconversion PL intensity power series, revealing a power law with a slope of 1.4475.